\documentclass[aps,prb,groupedaddress, amsmath,amssymb,showpacs,eqsecnum,floatfix,twocolumn]{revtex4}
\usepackage{graphicx}

\newcommand{\bsigma}{\mbox{\boldmath$\sigma$}}
\newcommand{\bnabla}{\mbox{\boldmath$\nabla$}}
\begin{document}


\title{Emergence of Triplet Correlations in Superconductor/Half Metallic
Nanojunctions with Spin Active Interfaces}


\author{Klaus Halterman}
\email{klaus.halterman@navy.mil}
\affiliation{Research and Intelligence Department, Physics Division, Naval Air Warfare Center, China Lake, California 93555, USA}
\author{Oriol T. Valls}
\email{otvalls@umn.edu}
\altaffiliation{Also at Minnesota Supercomputer Institute, University of Minnesota,
Minneapolis, Minnesota 55455}
\affiliation{School of Physics and Astronomy,
University of Minnesota, Minneapolis, Minnesota 55455, USA}

\date{\today}

\begin{abstract}
We study triplet pairing correlations induced in an SFS trilayer (where F is a 
ferromagnet and S an ordinary $s$-wave superconductor) by spin flip
scattering at the interfaces. We derive and solve self consistently the 
appropriate Bogoliubov-de Gennes equations in the clean limit. We find
that the spin flip scattering generates $m=\pm 1$  triplet correlations,
odd in time. We study the general spatial behavior of these and of $m=0$ correlations
as a function of position and of spin-flip strength, $H_{spin}$. We concentrate on the case 
where the ferromagnet is half-metallic. We find that for certain values of $H_{spin}$, the triplet correlations 
pervade the magnetic layer and can penetrate deeply into the superconductor.
The behavior we find depends very strongly on whether the singlet order parameter is 
in the 0 or $\pi$ state, which must in turn be determined self-consistently.  
We also present results for the density of states (DOS)
and for the local magnetization, which, due to spin-flip processes, is not in general aligned
with the magnetization of the half metal, and near the interfaces, rotates as a function 
of position and $H_{spin}$. 
The average DOS in both F and S is shown to exhibit various subgap bound states positioned
at energies that 
depend strongly on the particular junction state and the spin flip scattering strength.

\end{abstract}


\maketitle

\section{Introduction}
\label{intro}
 
Nanoscale structures
involving  Ferromagnet (F) and Superconductor (S) junctions 
illustrate the unique interaction of superconducting and ferromagnetic symmetries and
provides a novel opportunity to study  
the powerful influence that the spin degree of freedom plays in transport and thermodynamic 
properties of such systems.
The now well established variety
of phenomena\cite{buzdin}  induced by
the resulting proximity effects, 
includes exotic singlet superconducting 
correlations, in particular the
damped oscillatory Cooper pair amplitude in the magnet, 
with  a spatial decay length in the clean limit over a few nanometers for strong magnets (such as
Ni, Co, and Fe), and considerably less than
the superconductor coherence length $\xi_0$.
These oscillations lead to the possibility of switching between
0 and $\pi$ junction states, with considerable potential\cite{igor} for
applications. 
The superconductor region 
correspondingly becomes
affected as it
experiences the pair-breaking effects of the ferromagnet
and becomes locally magnetized.
These mutual effects depend considerably on the strength  
of the magnet and transparency of the interfaces, usually assumed spin-independent.
If the interface scattering is generalized to include spin dependence, 
where the spin of the impinging electron is flipped
when traversing the corresponding interface,
the whole picture can be modified, including the emergence or enhancement of exotic triplet states, that involve odd frequency
or different time triplet correlations.

Triplet pairing correlations can arise in ferromagnet and superconductor heterostructures involving
superconductors with a rotationally symmetric pairing symmetry ($s$-wave pairing), 
since they involve odd time symmetry pairing, as originally
proposed\cite{brz} in a different context. In F-S structures 
the proximity effects associated with the magnet break spin
rotation invariance.
The superconducting order parameter in
this scenario changes sign under time coordinate interchange of
the two electrons comprising Cooper pairs.
When a single quantization axis exists for the system, the only possible triplet pairing state 
is the one comprised of opposite spin pairs 
(the $m=0$ projection on the given quantization axis).
If there exists non-collinear or inhomogeneous magnetization in the system, as can occur 
in structures involving differently oriented F layers 
or 
an in-plane spin flip scattering potential, 
equal spin $m =\pm1$ triplet correlations can also arise.
Several investigations into triplet effects in superconductor and
ferromagnet hybrids has revealed a host of interesting and exotic phenomena,\cite{bergeret,bergeret2,keizer} including
the possibility of a long-ranged superconductivity 
proximity effect in F-S structures. 
A new superconducting state that can potentially extend
superconducting correlations into the magnetic region over long distances
brings with it a host of useful device application involving
low temperature nanodevices,
including nanoelectromechanical systems (NEMS), and superconducting circuits ($\pi$ junctions with $I_0<0$).

While there has been recently considerable interest in trying to isolate and detect 
the triplet pairing state
that is predicted to exist in such S-F structures,
it can  be difficult to disentangle the triplet and singlet correlations.
It is therefore of interest to investigate  heterostructures that  
restrict
the singlet order parameter somewhat, yet retain the
desired triplet correlations.
The pinpointing of triplet effects
can be exploited with the use of highly polarized materials, namely half metallic ferromagnets,
where only a single spin channel is present at the Fermi level.
The ordinary singlet pair amplitude is thus suppressed, since
the magnet behaves essentially as
an insulator for the opposite spin band. 
Several half metallic candidates are considered in connection with superconducting hybrids and spintronic applications.
These include the conducting ferromagnet\cite{keizer}  ${\rm CrO_2}$, the manganese perovskite\cite{cuoco,zalk}
${\rm La_{2/3} Ca_{1/3} Mn O_3}$, and the Heusler alloys,\cite{kubota}
possibly ${\rm Co_2 Fe Si}$ and ${\rm Co_2 Mn Si}$, which  are 
attractive from a nanofabrication standpoint, since  growth by sputtering techniques is applicable. 
This is pertinent since 
${\rm CrO_2}$ cannot be grown by sputtering and is metastable. 
Thus it is of interest to determine the circumstances under which
triplet correlations and related single particle signatures emerge when
a wide variety of spin flipping strengths for a half metallic
ferromagnet in F-S nanojunctions, with S being a conventional $s$-wave 
superconductor and interfacial spin flip scattering providing
the required symmetry breaking.

The spin flip processes and their participation in proximity effects have been  explored in
several different experimental setups\cite{velez,keizer,kriv,sang} 
possible in many instances due to advanced e-beam lithography and sputtering techniques. 
There is no current experiment however, 
that offers 
completely indisputable evidence of triplet correlations in ferromagnet and conventional superconductor hybrids,
and thus further experimental investigations are needed. 
The spin flipping associated with the intrinsic exchange field in the ferromagnet in S-F bilayers
was linked to critical temperature variations\cite{velez} as a function of F layer thicknesses.
It was suggested that the measured Josephson current\cite{keizer} in a sample with
two ${\rm NbTiN}$ ($s$-wave) superconductors coupled
by half metallic  ${\rm CrO_2}$   is due to a supercurrent
carried by spin triplet pairs since the electronic transport in ${\rm CrO_2}$ is metallic solely for the spin up band, and
the expected magnet thickness for that system exceeded the estimated singlet correlation length for that sample.
Current-voltage measurements,\cite{kriv} in singlet superconductor-half metallic point contacts revealed a marked resistance
decrease and increased normal state conductance at small voltages, attributed to spin singlet-triplet conversion.
Single particle spectroscopy results for 
the density of states (DOS) were also reported\cite{sang} for F-S bilayers 
including the strong ferromagnet, Ni. 
In that experiment, the conductance signature was measured as a function of ferromagnet thickness, $d_F$,
revealing an interesting double peak structure and other subgap features that could not be theoretically accounted for
within the dirty limit framework.

Numerous theoretical approaches involving spin dependent scattering of some sort have helped pave the way 
towards unveiling the role of triplet pairing correlations in diffusive SFS hybrid nanostructures
or clean SFS junctions, both within the quasiclassical\cite{eschrig} regime.
The intermediate regime, separating diffusive and ballistic motion was also studied quasiclassically.\cite{linder3}
Any purely microscopic approaches, that retain
quasiparticle information at the atomic scale, typically involved spin independent scattering potentials
at the interfaces.\cite{hvb,bvh}
If the pair-breaking mechanism of spin-flip 
scattering at the interfaces is included, the resultant interchange of spins yields  
complicated normal and Andreev reflection events.
The investigation into these issues have been predominately in the 
diffusive regime however.
For instance, with relatively thick half metals, Josephson coupling can occur via triplet correlations from 
the singlet superconductor and spin mixing occurring at a spin active interface.\cite{eschrig}
For calculations neglecting the 
mutual influence of superconducting and ferromagnet order parameters, 
spin flip scattering was shown to also have a detrimental effect on the residual supercurrent,
thus limiting such junctions as useful spin switches.\cite{linder2}
A long range triplet component can arise\cite{fominov} when the ferromagnet has a N{\'e}el domain structure where the in-plane magnetization
rotates with changing depth in the magnet.
The odd frequency pairs arising from spin flipping at the junction interfaces
can cause a peak in the local DOS of a diffusive half metallic ferromagnet.\cite{asano}
If the ferromagnet can be modeled by a conical magnetic structure,\cite{annett} as in Holmium, 
it was found that both singlet and triplet correlations undergo short range decay. 
Also, the decay length of the Josephson current was shown to decrease\cite{faure} with spin-flip and spin-orbit scattering,
with spin-orbit scattering typically being the more destructive of the two.
By illuminating the junction with microwave radiation at the proper resonance frequency however, the
critical current can be enhanced,\cite{tak} due in part to singlet-triplet conversion processes. 
Nearly all of the cases studied thus far involve the quasiclassical method,
and it is unclear how this landscape is modified when atomic length scales
are not eliminated in the pertinent equations and when self consistency of the singlet order parameter is taken into account.

In this paper, we 
address some of the above issues by presenting 
a fully self consistent framework for a clean nanoscale trilayer junction comprised of
a half metal sandwiched between two conventional, $s$-wave superconductors.
The  pair breaking mechanism is spin flip scattering at the interfaces, which produces
$m=\pm 1$ odd time pairs, and modifies the triplet $m=0$ component.
The presence of two S layers, coupled through F via the proximity effect allows us
to compare and contrast the 0 and $\pi$ states.
Our method is based on the quantum mechanical Bogoliubov-de Gennes (BdG)
equations in the clean limit, which is
ideal 
for half metallic ferromagnets and proximity effects that can involve
singlet correlations in the magnet with very short decay lengths of just a few nanometers.
We are able to fully take into
account proximity effects in the magnet and ``inverse proximity effects" that arise 
in the superconductor regions, including the presence of a magnetic moment
component normal to the magnetization in F.
We employ a recently developed\cite{kvb} method to determine the
triplet correlations in such structures using a Heisenberg representation to 
derive the time dependent quasiparticle wavefunctions.
We consider spin-active interfaces by incorporating 
in-plane spin dependent scattering in the effective Hamiltonian.
By varying the spin scattering strength parameter, $H_{spin}$,
over a broad range, long range triplet correlations
are shown to emerge and evolve. We study the spatial profile
of all possible triplet correlations which depends on their 
corresponding projection onto the axis
of quantization, which is taken to be along the fixed direction
of magnetization in the half metal (the $z$ axis in our case).
The relative admixtures of triplet amplitudes with
total spin projection $m=0$ on
the $z$-axis (labeled $f_0$) and those
with $m=1$ total spin projection quantum number, $f_1$,
depend
crucially on whether
the junction is in a $0$ or $\pi$ state.
The junction state with the lowest free energy, and its corresponding stability, is dictated not only 
by the geometry and spin splitting strength of the magnet,\cite{stab} 
but also by the magnitude of spin scattering at the interfaces. 
An accurate determination of this requires a self consistent calculation: this is 
even more evident when one considers that
the triplet correlations in general peak near the interfaces,
where self-consistency is most critical.
After presenting the triplet correlations within the system as a function of the spin scattering strength, 
we turn our attention to
the effect spin active interfaces have on other physically important single-particle
quantities such as the average DOS, and local magnetic moment. We find that the DOS has a subgap signature that
depends on whether the junction is in a $0$ or $\pi$ state and the degree of spin scattering at the interfaces.

The rest of this paper is organized as follows: in Sec.~\ref{methods} we discuss the methods we use to 
evaluate the order parameter in a fully self consistent way and to determine the triplet 
correlations. As pointed out in Ref.~\onlinecite{hvb}, self consistency is absolutely essential in order to 
correctly obtain the odd parity triplet correlations: without self consistency the Pauli 
principle is violated. We review and discuss the appropriate quasiparticle expansions, the evaluation of the  
matrix elements, and other relevant details
in the solution of the corresponding eigenvalue problem.
The definition of the time-dependent triplet amplitudes is given and other
quantities that are also of interest, such as the local magnetic moment and the local 
DOS are defined. Then, in Sec.~\ref{results}, the condensation energy, which reveals the
relative stability of the 0 and $\pi$ phases, is calculated as a function of $H_{spin}$. We then 
discuss in detail our results for the spatial and time behavior of the 
triplet
amplitudes  as a function of the spin flip 
scattering strength. The associated penetration depths of the equal spin triplet amplitudes into the superconductor
reveal the long range nature of these correlations. Next, we  focus on the averaged 
DOS in each region of competing
order parameter symmetries, and display 
the spatial dependence of the local magnetic moment, 
as it relates to the inverse proximity effect. 
We show that the induced magnetic moment vector rotates near the interfaces. 
Finally, we give a summary of our results in Sec.~\ref{conclusions}.

\section{Methods}
\label{methods}

The SFS junction that we study is a trilayer structure  infinite in the plane
parallel to the interfaces, which we label the $x-z$ plane,
and with total length $d$ in the $y$ direction, normal
to the interfaces. The width of each of the two superconductor layers
is labeled by $d_S$ and that of the ferromagnet
by $d_F$. The superconductors are $s$-wave and identical. The entire structure
occupies the space $0\leq y \leq d$, with one superconductor occupying,  $0 \leq y \leq d_S$, 
the ferromagnet: $d_S \leq y \leq d_S+d_F$, and the other superconductor, 
$d_S+d_F < y \leq d$. There is spin-flip scattering at the interfaces, which
will be described below.

To determine the equations governing
triplet effects in SFS nanojunctions, we start with
a fundamental quantity, the effective Hamiltonian, ${\cal H}_{\rm eff}$,
written in terms of creation and annihilation field operators and vector Pauli  
spin matrices, $\bsigma$,
\begin{widetext}
\begin{align}
{\cal H}_{\rm eff} = &\int d^3 r\Bigl\lbrace \sum_\alpha \psi^\dagger_\alpha({\bf r})  
 \left[-\frac{\bnabla^2}{2m} -E_F+V_0({\bf r})\right]  \psi_\alpha({\bf r})
+\sum_{\alpha,\beta} \psi^\dagger_\alpha({\bf r})({\bf V} \cdot \bsigma)_{\alpha \beta}\,\psi_\beta({\bf r}) 
 \nonumber \\
&+\frac{1}{2} [\sum_{\alpha,\beta}  (i\sigma_y)_{\alpha \beta} \Delta({\bf r}) \psi^\dagger_\alpha({\bf r}) \psi^\dagger_\beta({\bf r})+ \rm{h.c.}] 
+\sum_{\alpha,\beta} \psi^\dagger_\alpha({\bf r})({\bf h} \cdot \bsigma)_{\alpha \beta}\,\psi_\beta({\bf r}) \Bigr\rbrace.
\end{align}
\end{widetext}
The first term in brackets is the single particle Hamiltonian for a 
quasiparticle with effective mass, $m$, Fermi energy, $E_F$, and scattering from a spin independent potential $V_0({\bf r})$.
The pair potential, $\Delta ({\bf r})$,
characterizes the spatial dependence to the superconducting singlet correlations, 
and will be calculated in a self consistent fashion as described below.
The ferromagnetic exchange 
field, ${\bf h}(y)=h_0 \hat{\bf z}$, representing the ferromagnetism, is taken as constant in the F layer and
vanishing in the two S layers, and it is along the $\hat{\bf z}$ axis of quantization.
This intrinsic exchange field in the magnet, 
favoring a given spin, thus contributes to the overall behavior of triplet correlations.
The important spin flip scattering 
will be assumed to be confined to the two interfaces near $y=d_S$ and $y=d_S+d_F$. 
It takes
place in the invariant $x-z$ plane:
${\bf V} \cdot \bsigma=V_x(y)\sigma_x+V_z(y)\sigma_z$.
Its $z$ component represents a less important local modification of the $h_0$ 
field, while $V_x$ is the spin flip term. 
We have taken $V_y=0$  because of the geometry and also for convenience: including $\sigma_y$
terms precludes the use
of exclusively real numbers in the numerical diagonalizations and leads
to additional technical irrelevant complications.
Each of the triplet states  can potentially exist over large length scales,
thus allowing  competing orderings to coexist.

To solve the problem
we 
diagonalize ${\cal H}_{\rm eff}$ via a Bogoliubov transformation. The details
are given elsewhere\cite{hvb} and need not be repeated here. 
Through the
use of  standard commutation relations, we end up after some straightforward algebra,
a general  
coupled four component set of equations. This leads to a generalization 
of the textbook\cite{bdg} Bogolioubov-de Gennes (BdG) equations, which 
give rise ultimately to spin singlet and triplet amplitudes.
By making use of the Pauli spin matrices and of a set of Pauli-like 
matrices \mbox{\boldmath$\rho$} 
in particle-hole space, 
the general time and spin-dependent BdG equations can be expressed compactly as,
\begin{equation}
\left[
\rho_z\otimes\left({\cal H}_0 \hat{\bf 1} -(h_z-V_z)\sigma_z\right)+\left(\Delta(y)\rho_x
+V_x \hat{\bf 1}\right)\otimes \sigma_x \right]{\Phi}_n(y,t)=i\frac{\partial {\Phi}_n(y,t)}{\partial t},
\label{bogo}
\end{equation}
where the four component wavefunction,
${\Phi}_n(y,t)$, is
a vector of quasiparticle amplitudes,
${\Phi}_n(y,t)\equiv(u_{n\uparrow}(y),u_{n\downarrow}(y),v_{n\uparrow}(y),v_{n\downarrow}(y))^{\rm T} e^{-i
\epsilon_n t}$,
where the superindex
denotes transposing, the $u_{n \sigma}$  and $v_{n \sigma}$ have their standard\cite{bdg} meaning 
as quasiparticle amplitudes 
and $\epsilon_n$ is the eigenenergy. We have assumed here  that $E_F$  is the same throughout
the sample: the majority ($+$) and minority ($-$) bandwidths in F are $E_F \pm h_0$. 
The single particle quasi one-dimensional Hamiltonian ${\cal H}_0$ becomes in our geometry
\begin{equation}
{\cal H}_0\equiv \frac{1}{2m} \frac{\partial^2}{\partial y^2}
+\varepsilon_{\perp} 
-E_F+ V_0(y),
\end{equation}
where
$\varepsilon_{\perp}$ is  the 
energy in the transverse direction.
Carrying the time derivative through, and taking the outer product in Eq.~(\ref{bogo}), we 
can rewrite Eq.~(\ref{bogo}) in the much less compact but intuitively 
more immediate form:
\begin{widetext}
\begin{align}
&\begin{pmatrix} 
{\cal H}_0 -h_z(y)+V_z(y)&V_x(y)&0&\Delta(y) \\
V_x(y)&{\cal H}_0 +h_z(y)-V_z(y)&\Delta(y)&0 \\
0&\Delta(y)&-({\cal H}_0 -h_z(y)+V_z(y))&V_x(y) \\
\Delta(y)&0&V_x(y)&-({\cal H}_0+h_z(y)-V_z(y)) \\
\end{pmatrix} \nonumber \\ 
&\hspace{3.5in} \times
\begin{pmatrix} 
u_{n\uparrow}(y)\\u_{n\downarrow}(y)\\v_{n\uparrow}(y)\\v_{n\downarrow}(y)
\end{pmatrix}
=\epsilon_n
\begin{pmatrix}
u_{n\uparrow}(y)\\u_{n\downarrow}(y)\\v_{n\uparrow}(y)\\v_{n\downarrow}(y)
\end{pmatrix}\label{bogo2},
\end{align}
\end{widetext}
where
the spin dependent interface scattering potential 
should be understood to be given  in terms of delta function scatterers:
$V_i(y) = V_i [\delta(y-d_S) + \delta(y-(d_S+d_F))]$, and as explained above,  $i=x,z$. 
The convenient dimensionless parameter $H_{spin} \equiv 2 m V_x/k_F$ characterizes the strength of  
the interface scattering.
The  spin flip $x$-component, $V_x(y)$, technically complicates the calculation and prevents 
the simple splitting of the BdG equations into two separate equations by means of symmetry relations,
as in the case of collinear magnetizations, or when a single quantization axis exists for the whole system.
Here, all four components are needed since the
exchange field in the ferromagnet as well as the spin-flip potential break the spin rotation invariance.

The general expression for the self consistent pair potential, valid for
all temperatures, $T$, is given by,
\begin{equation}  
\label{del} 
\Delta(y) = \frac{g(y)}{2}{\sum_n}    
\bigl[u_n^\uparrow(y)v^\downarrow_n (y)+
u_n^\downarrow(y)v^\uparrow_n (y)\bigr]\tanh\Bigl(\frac{\epsilon_n}{2T}\Bigr), \,
\end{equation} 
where the sum is over eigenstates (the index $n$ now subsumes  not only the  
quantized index in Eq.~(\ref{bogo2}) but also the transverse energies
$\varepsilon_\perp$) which is performed  over all eigenstates with 
positive energies smaller than or equal to  the ``Debye'' 
characteristic energy cutoff $\omega_D$, and  
$g(y)$ is the superconducting coupling parameter that is a constant 
$g_0$ in the intrinsically superconducting
regions and zero elsewhere. 

The triplet correlation functions, odd in time, which are the main subject 
of  our study are
defined\cite{hvb} in terms of the usual field operators as, 
\begin{subequations}
\label{alltriplet}
\begin{align}
f_0 ({\bf r},t) \equiv \frac{1}{2}[\langle \psi_\uparrow({\bf r},t)\psi_\downarrow({\bf r},0) \rangle+
\langle \psi_\downarrow({\bf r},t)\psi_\uparrow({\bf r},0) \rangle]
\label{f0def}
\\
f_1 ({\bf r},t) \equiv \frac{1}{2}[\langle \psi_\uparrow({\bf r},t)\psi_\uparrow({\bf r},0) \rangle-
\langle \psi_\downarrow({\bf r},t)\psi_\downarrow({\bf r},0) \rangle],
\label{f1def}
\end{align}
\end{subequations}
where we are clearly free to choose one time coordinate to be zero, without loss of
generality. These correlation functions must be odd in time because
of the Pauli principle. Hence they vanish identically at $t=0$. At $T=0$, these 
expressions are conveniently written
in terms of the quasiparticle amplitudes\cite{kvb}:
\begin{subequations}
\label{alltripleta}
\begin{align}
f_0 (y,t) & = \frac{1}{2} \sum_n \left[ u_{n\uparrow} (y) v_{n\downarrow}(y)-
u_{n\downarrow}(y) v_{n\uparrow} (y) \right] e^{- i \epsilon_n t}, 
\label{f0defa}
\\
f_1 (y,t) & = \frac{1}{2} \sum_n \left[ u_{n\uparrow} (y) v_{n\uparrow}(y)+
u_{n\downarrow}(y) v_{n\downarrow} (y) \right] e^{- i \epsilon_n t},
\label{f1defa}
\end{align}
\end{subequations}
where {\it all} positive energy states are in general summed over.
In practice,  
we find that at
finite times, results become cutoff independent beyond a
value a few $\omega_D$. However, 
to ensure the vanishing of the triplet components at $t=0$,  it is
necessary to sum over 
a much larger energy range.

Besides the pair potential
and the triplet amplitudes, we can also determine various physically relevant single-particle quantities.
One such important quantity is the local magnetization, which is a measure
of the so-called inverse proximity effect, and can be particularly useful in 
characterizing the 
magnetizing effects in the superconductor as a result of the localized spin flip interface scattering and
intrinsic exchange field of the magnet. It can also serve as 
an effective self-consistent measure of the magnetization field in the half metal.
Recent magneto-optical Kerr effect measurements\cite{xia} of a superconductor/ferromagnet bilayer revealed 
that the superconductor 
became magnetized, illustrating the need to determine the spatial behavior of the magnetization fully.
In the presence of spin flip scattering, the local magnetic moment ${\bf m}$ will depend on
the coordinate $y$ (in our geometry) and, in the presence of the spin flip term it will
have in general both $x$ and $z$ components, ${\bf m} = (m_x,0,m_z)$.
In terms of the quasiparticle amplitudes calculated from the self-consistent BdG equations 
we have,
\begin{subequations}
\label{mm}
\begin{eqnarray}
m_z(y) = -\mu_B \sum_n [v_{n\uparrow}^2(y)-v_{n\downarrow}^2(y)], \\
m_x(y) = -2\mu_B \sum_n v_{n \uparrow}(y) v_{n \downarrow}(y),
\end{eqnarray} 
\end{subequations}
where $\mu_B$ is the Bohr magneton.
The sums in Eqs.~(\ref{mm})  involve a sum over eigenstates, 
as in Eqs.~(\ref{del}) 
and (\ref{alltripleta}), although the energies
$\epsilon_n$ do not now appear explicitly. 

A very useful  tool in the study of these phenomena is experimental tunneling experiments,
where spectroscopic information, measured ideally by an STM, can reveal the local DOS. Therefore
we have computed here also the local DOS $N(y,\epsilon)$ as a function of $y$. We have  
$N(y,\epsilon) \equiv N_\uparrow (y,\epsilon) + N_\downarrow (y,\epsilon)$, where,
\begin{align}
\label{dos}
N_\sigma(y,\epsilon) = \sum_n [u_{n\sigma}^2(y) \delta(\epsilon-\epsilon_n)+ v_{n\sigma}^2(y) \delta(\epsilon+\epsilon_n)], 
\quad \sigma = \uparrow, \downarrow. 
\end{align} 



In order to numerically solve the problem, we must re-express the equations in terms of 
matrix elements in an appropriate basis.
These matrix elements are obtained via projection upon a orthonormal complete set,
that inherently satisfies the boundary conditions 
of vanishing wavefunction at the outer edges of the trilayer structure. Thus we write 
$u_{n \alpha}(y) = \sqrt{2/d} \sum_{q=1}^N u^\alpha_{n q} \sin(q\pi y/d)$, 
$v_{n \alpha}(y) = \sqrt{2/d} \sum_{q=1}^N v^\alpha_{n q} \sin(q\pi y/d)$,
with $\sigma = \uparrow, \downarrow$. Inserting these into Eq.~(\ref{bogo2}), we have
the general $4N\times 4N$ matrix consisting of a $4\times 4$ array of block submatrices, each
of rank $N$:
\begin{align}
&\begin{pmatrix} 
{\cal H}^+ &{\cal V}^x &0&\cal{D} \\
{\cal V}^x&{\cal H}^-&\cal{D}&0 \\
0&\cal{D}&-{\cal H}^+&{\cal V}^x \\
\cal{D}&0&{\cal V}^x&-{\cal H}^- \\
\end{pmatrix} 
\Psi_n=\widetilde{\epsilon}_n
\Psi_n,
\label{bogo3}
\end{align}
where we measure all
energies in terms of
$E_F$, so that $\widetilde{\epsilon}_n \equiv \epsilon_n/E_F$, $Z_{Bz}\equiv m V_z/(k_F^2 d)$,
and the vector $\Psi_n$ is 
the transpose of $(u^\uparrow_{n 1}, \cdots , u^\uparrow_{n N},  u^\downarrow_{n 1}, \cdots , u^\downarrow_{n N},
v^\uparrow_{n 1} , \cdots, v^\uparrow_{n N}, v^\downarrow_{n 1}  \cdots ,v^\downarrow_{n N})$. We
find, after lengthy but elementary algebra the matrix elements:
\begin{widetext}
\begin{align}
{\cal H}^{\pm}_{i j} & = \Biggl[ \Bigl(\frac{i \pi}{k_F d}\Bigr)^2+\frac{\varepsilon_\perp}{E_F} -1 \mp 
I\Bigl[\frac{d_F}{d}+K^{(1)}_{2i}-K^{(2)}_{2i}\Bigr] 
\pm Z_{Bz}(U^{(1)}_i U^{(1)}_j+U^{(2)}_i U^{(2)}_j)\Biggr]\delta_{ij} \nonumber \\ 
&  \mp I[K^{(1)}_{i+j}-K^{(1)}_{i-j}+K^{(2)}_{i-j}-K^{(2)}_{i+j}] \pm Z_{Bz}(U^{(1)}_{i}U^{(1)}_{j}+U^{(2)}_iU^{(2)}_j), 
\end{align}
\end{widetext}
where
$I \equiv h_0/E_F$ and $Z_{Bz}\equiv 2V_z/E_Fd$. 
The important  spin flip component, off the main diagonal, giving rise to equal spin
triplet correlations is,
\begin{equation}
{\cal V}^x_{ij} = Z_{Bx}[U^{(1)}_i U^{(1)}_j+U^{(2)}_i U^{(2)}_j],
\end{equation}
where $U^{(1)}_q = \sin(q\pi d_S/d)$, $U^{(2)}_q = \sin(q\pi (d_F+d_S)/d)$, $K^{(1)}_q=U^{(1)}_q/(q\pi)$, 
$K^{(2)}_q=U^{(2)}_q/(q\pi)$, and  $Z_{Bx}$ which arises 
from the spin-flip scattering is $Z_{Bx} \equiv 2 V_x/E_F d$. 
This is related to the parameter $H_{spin}=2mV_x/k_F $ defined earlier to characterize the spin flip 
scattering, by $H_{spin}=k_F d Z_{Bx}/2$. 
Without loss of generality we
can take $V_z=V_x=V_{spin}$. 
The matrix elements corresponding to the pair potential are, 
${\cal D}_{ij} = {2}/(E_F d) \int_0^d dy \Delta(y) \sin(i\pi y/d) \sin(j\pi y/d)$, 
recalling that $\Delta(y)$ vanishes in the magnet layer, due to the coupling $g(y)$.
Since we are not permitted to use previous symmetry relations among the quasiparticle 
amplitudes and energies that reduced the matrix eigensystem to $2N$, 
we are forced, as mentioned above, to solving the $4N\times4N$ system,
and retaining only the positive energy states. As in previous work, the diagonalization
is performed iteratively until the self-consistency condition Eq.~(\ref{del}) is satisfied.



\section{Results}
\label{results}

The results of our calculations are described in detail in this section.
We will measure all the lengths in units of the Fermi wave vector $k_{F}$, 
and define the relative dimensionless coordinate $Y\equiv
k_{F}(y-d/2)$, i.e. $Y=0$ is at the center of the junction. 
All  times will be given in units of $\omega_D^{-1}$ via
the dimensionless time $\tau \equiv \omega_D t$.
In the ferromagnet we have for the spin up and spin down band widths, 
$E_\uparrow=E_{F}+h$ and $E_\downarrow=E_{F}-h$. 
The dimensionless 
measure of the intrinsic exchange energy in the magnet is
$I\equiv h/E_{F}$. All results below are for the
half-metallic limit, $I=1$, discussed recently in the context of spintronic
materials.\cite{groot} The spin flip scattering at the interface is
characterized, as previously explained, by the dimensionless parameter
$H_{spin}$ for which we will consider values between zero and unity.
Geometrically, we will consider a system consisting of two  thick
superconducting layers, each  of a thickness $d_S$ such that 
$D_S \equiv k_{F}d_S =300$. Since odd-frequency or different-time triplet
states arise from 
magnetic effects at the interfaces and exchange field in the half metal,
this broad range of parameters will give revealing hints as to their existence.
The chosen $d_S$ 
considerably exceeds the superconducting coherence length $\xi_0$, which
we take to be $\xi_0=50 k_F^{-1}$. Thus 
$d_S= 6 \xi_0$, so that
we can disentangle quantum interference effects from the modified Andreev and
scattering events at a spin flip interface.
The two S layers are separated by
a ferromagnetic layer, which must be taken to be thin enough so that the two
superconductors are still coupled through the F material via the proximity
effect. We take $D_F \equiv k_{F} d_F=10$. All results are computed in 
the low temperature limit.

\begin{figure}
\includegraphics[width=3.4in]{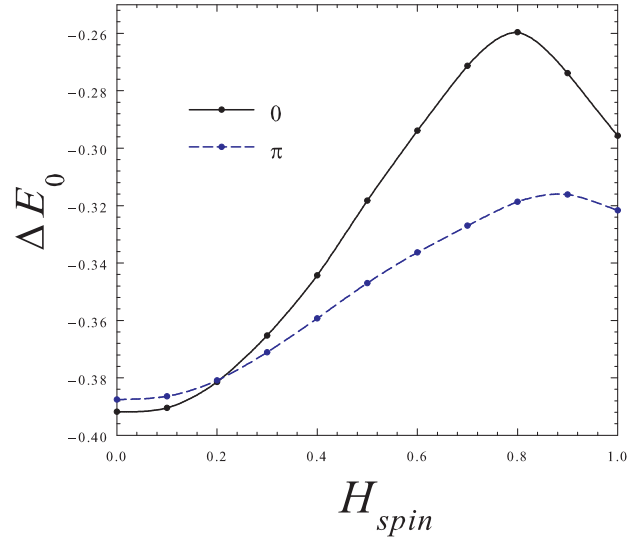}
\caption{(Color online)
The dimensionless condensation energy (free energy at $T=0$ in units
of $N(0)\Delta_0$, see text),
versus the
dimensionless parameter $H_{spin}$ characterizing the spin flip strength
at the interface, for an SFS junction with a half metallic ferromagnet. For the geometry chosen, we see that
the both the $\pi$ 
and the $0$ states are stable for all values of $H_{spin}$ considered
and that the $\pi$ state has the lower condensation energy except at  small
values of $H_{spin}$. }
\label{fig1}
\end{figure} 

Each junction between
two consecutive S layers can be of the ``0" type
(with the order parameter in both S layers having the
same sign) or of the ``$\pi$" type (opposite sign). 
The
characteristics of a $0$ or $\pi$ junction are directly
connected to the spatial behavior of the pair potential $\Delta(y)$ 
and,  to determine its precise form, this quantity must be calculated 
self-consistently so that the resulting singlet pair amplitude corresponds
to a minimum in the free energy. 
The relative stability of the different
states that may be obtained through self-consistent solution
of the BdG equations 
depends on the free energy of the junctions. 
We therefore consider first the stability of the system for 
our parameter values and
geometry.
In Fig.~\ref{fig1}  we plot the condensation energy (the free energy at $T=0$)
of the system in dimensionless form, that is, in units of $N(0) \Delta_0$ where
$N(0)$ is the usual single spin density of states in the S material, and
$\Delta_0$ the bulk value of the gap. Thus, the quantity plotted would be $-1/2$
for a bulk S sample. This energy is calculated 
with high precision as explained in previous\cite{bvh} work. 
We see in the figure that,  
consistent with the results of Ref.~\onlinecite{bvh}, the condensation 
energies are  reduced, in all cases, from what they would
be for a bulk S sample. For all values of $H_{spin}$
both the 0 and $\pi$ configurations of the structure are at least locally
stable, but in general
non-degenerate, showing that indeed the two S slabs are indeed  
coupled via the 
proximity
effect. At very
small values of  the spin flip 
parameter, the $0$ configuration is  the  stable one,
but this changes as 
$H_{spin}$ increases: there is a first order phase
transition at $H_{spin} \lesssim 0.2$ and  
at larger values of $H_{spin}$ the $\pi$ configuration is
the stable one and the 0 configuration is
much less stable. 
The 0 configuration metastable minimum is shallowest at $H_{spin} \approx 0.8$
where the condensation free energy has a sharp maximum. The condensation free
energy of the $\pi$ state is more weakly dependent on $H_{spin}$ with a 
maximum near $H_{spin}=0.82$ much shallower than that found in the 0 state. 
Note that by decreasing the width of S, the qualitative results remain, 
but the overall results are shifted towards zero, resulting
in the possible elimination of the zero state altogether. In general,
computational convergence time is increased as
the condensation for a given state approaches zero.

\begin{figure*}
\includegraphics[width=5.5in]{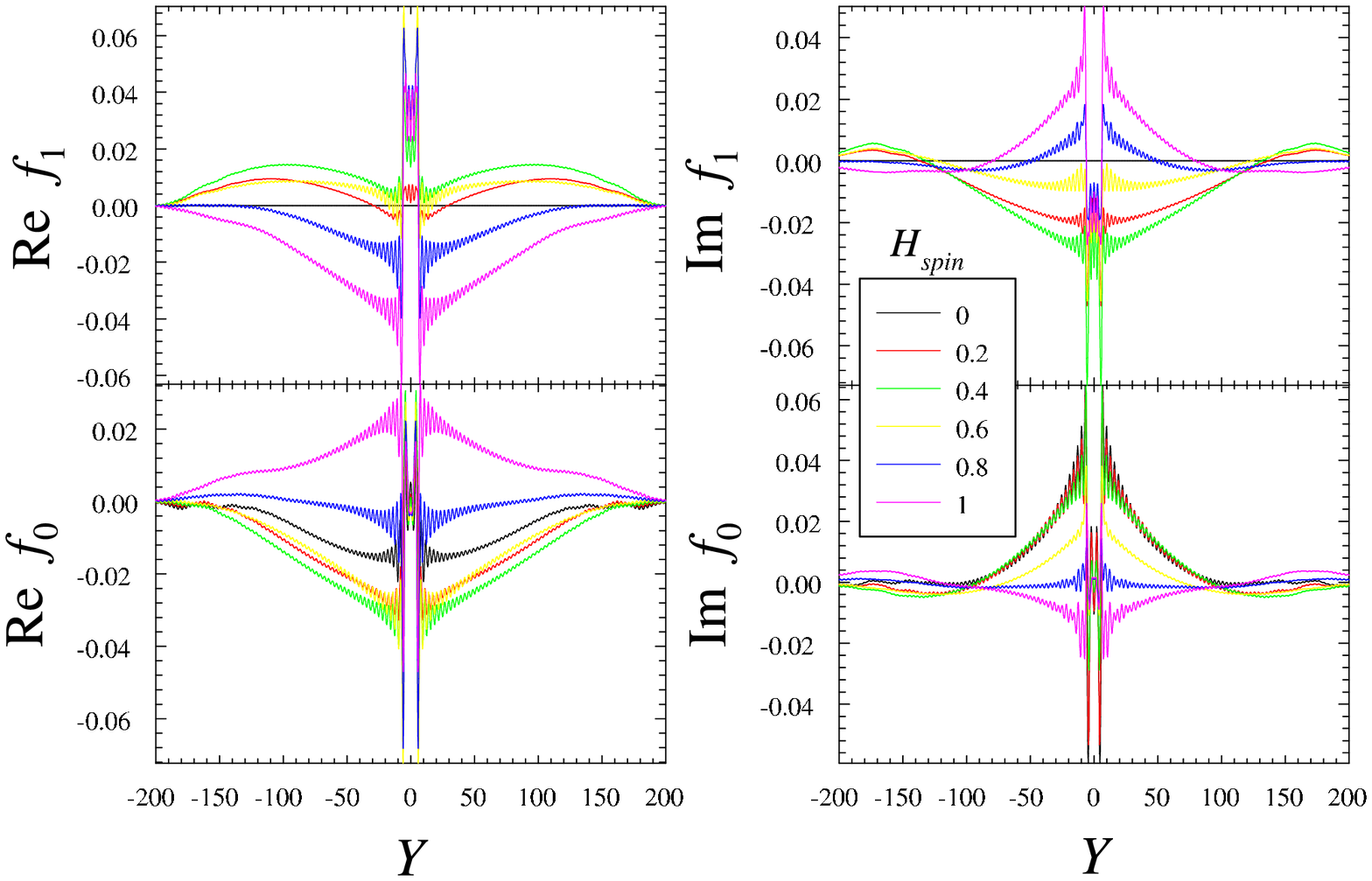}
\includegraphics[width=5.5in]{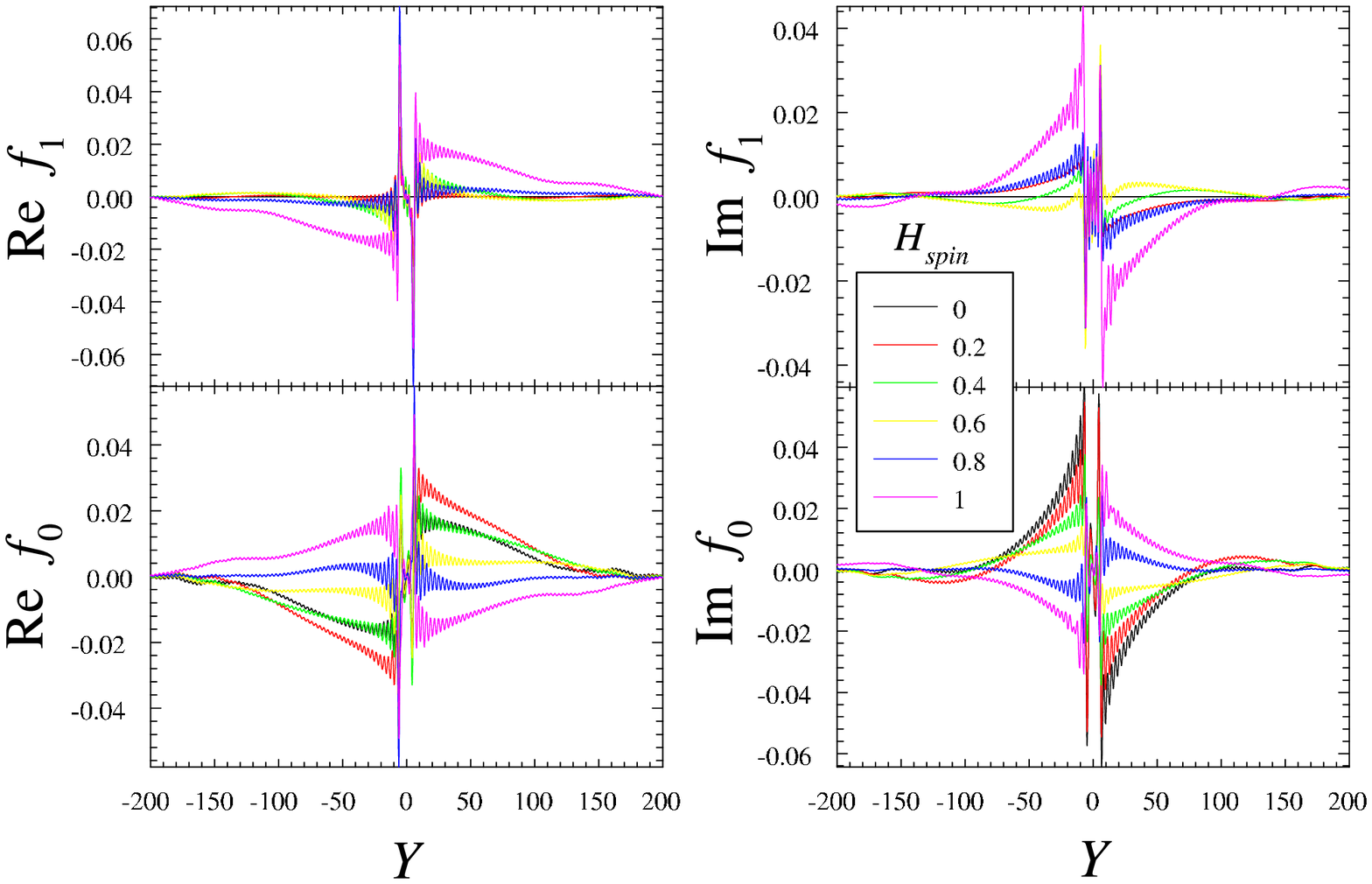}
\caption{(Color online)
The $f_1$ and $f_0$ triplet pair amplitudes (Eq.~(\ref{alltripleta})) 
for a $0$-junction (top 4 panels) and $\pi$-junction (bottom 4 panels)
plotted as a function
of the dimensionless coordinate $Y$ for several values of $H_{spin}$ as indicated in
the legend. The left panels show the real parts
while the right ones show the imaginary parts,
for values of $H_{spin}$ ranging from 0 to 1.
All results are at a fixed
value of the dimensionless time, $\tau=20$. }
\label{fig2}
\end{figure*}
We turn next to the spatial dependence of the
general complex triplet pairing functions $f_0(Y,\tau)$ and $f_1(Y,\tau)$ as
defined in Eqs.~(\ref{alltripleta}). 
In Fig.~\ref{fig2} we plot the corresponding
triplet amplitudes for each of the two types of solution (0 or $\pi$) 
as a function of the dimensionless coordinate $Y$ and at 
fixed time, $\tau=20$, for 6 equally spaced values of $H_{spin}$ 
in the range $0 \le H_{spin} \le 1$ (see legends). 
All amplitudes plotted are normalized to $\Delta_0/g$,
so that, if the similarly
normalized ordinary singlet amplitude were plotted, it would
reach unity deep in bulk S material. The value $\tau=20$ is chosen as
being near\cite{bvh} that which maximizes the 
correlations, 
and is such that the triplet pairing states
have penetrated most of the two superconductor regions. 
Results for the first group of 4 plots are for the $0$-state solutions, 
and with the real
and imaginary triplet amplitudes labeled accordingly.
The bottom series of four panels are for the $\pi$ junction counterparts.
The range of $Y$ included in the plots is, for clarity,
somewhat  narrower than the sample
size: regions where the amplitudes are very small or zero are omitted.
One can see that the amplitude $f_1$ vanishes identically  
in the absence of spin-flip scattering,
since in that case both the total spin and its $z$ component are good quantum
numbers. For finite values of the spin flip parameter, all possible projections
of the total spin exist.
The spatial symmetry of the singlet Cooper pair
is also reflected in the triplet pairing states:
it is evident from this and the next figure 
that if the singlet order parameter
is in a $0$ junction state, the corresponding
triplet amplitudes maintain that symmetry. This holds true for the spatially
antisymmetric $\pi$ junction results as well (bottom 4 panels).

Turning our attention to the real part of $f_1$ for
the  $0$-state we see that it 
shows
a monotonic decline in magnitude from the interface, over about 
three to four coherence lengths,  
for the largest two spin flip strengths 
(with a superposition of rapid oscillations).
The remaining weaker scattering strengths are quite different in that they yield 
nonmonotonic behavior with
a maximum deep into S at about $2\xi_0$, and then decaying to zero at 
roughly $4 \xi_0$ 
(hence for $H_{spin}=0.8$, this correlation dies about $\xi_0$ earlier). 
These amplitudes , ${\rm Re} f_1$, 
are predominately positive for higher spin transparency 
(smaller $H_{spin}$) junctions,
and then undergo a sign flip for the stronger $H_{spin}$.
If we examine now the imaginary component ${\rm Im} f_1$, 
still for the 0 state, 
we see similar opposite parity effects separating the strongest $H_{spin}$
from the weaker values, here however there is a clearer
separation between the curves. The time dependence here is noticeably different
than for the real part; the triplet correlations
have a faster rate of propagation, in that they have reached deeper within
the sample for the same $\tau$.
We have emphasized the triplet amplitudes in the S region, however
these plots reveal that besides the expected fact 
that $\langle \psi_\uparrow({y},t)\psi_\uparrow({y},0) \rangle$ is not 
destroyed
by the half metal, by including the proximity effects in a self consistent way,
we found  non-negligible different spin triplet pairing
in the half metal, but with a smaller magnitude that $f_1$.

The $\pi$ state results for the equal spin pairing correlations have 
markedly different profiles than those for the $0$ state: 
Besides being highly peaked at the interface, where spin-flip
scattering originates,  ${\rm Re}f_1$ has a very weak 
dependence on $H_{spin}$, with 
an abrupt emergence 
only for the  highest $H_{spin}$ and 
then still decaying over roughly 
the same distance in S. 
Overall, the $\pi$ state $f_1$ amplitudes are suppressed,
even for the case when both the $0$ and $\pi$ states have the
same condensation energy ($H_{spin}\approx 0.2$).
The diminished $\pi$ state results arises partly from the symmetry requirements 
imposed upon $f_i$: $f_i(-Y)=-f_i(Y)$,
thus in F, the $f_i$ amplitudes vanish at $Y=0$, which can constrain
the overall longer range spatial behavior.
It can be concluded that the singlet Cooper pair order parameter that minimizes the free energy,
and which $|\Delta(Y)|$ is typically larger,
does not necessarily result in the larger triplet amplitudes.
The imaginary parts can be discerned from the figure, but
clearly the imaginary parts to the amplitudes also do not simply differ
by a phase. Also, some amplitudes oscillate (with periods on the order 
of $\xi_0$)
while other plots show simple declines, with atomic scale oscillations 
superimposed on the general profile.

The $m=0$ triplet amplitude, $f_0$, with zero
projection of the $z$-component of total spin, 
does not vanish, at any finite time, even for $H_{spin}=0$ 
since the total
spin (as opposed to its $z$ component) is not a good quantum number in the
presence of the F material. It does still vanish at $t=0$ because of the Pauli
principle. Thus, in general, when a single quantization axis exists for the
system, $f_0$ coexists with the ordinary singlet $s$-wave component. 
We  expect the results for $f_0$ to be different those for $f_1$ since
$f_0$ does not emerge solely from $H_{spin}$: 
there are two competing spin flip effects in the $z$ direction:
the magnetization of the half metal, and the spin dependent scattering at the
interface. It is clear from the form of the BdG equations (\ref{bogo2}) 
that these 
two effects compete against each other.
For both the 0 and $\pi$ state,  
the absolute value of a given component of $f_0$ at fixed time in S
is again a non-monotonic function of $H_{spin}$, but the 
overall dependence 
is visibly different. 
The 
maximum value of $f_0$ in S is always at the interface: this is as expected from
the above considerations regarding 
quantum numbers. 
The
behavior with $H_{spin}$ echoes (at larger
values) 
that found for $f_1$, and for the same reasons.
In all cases,
these triplet correlations clearly pervade the thin F layer,
while their penetration into S increases only weakly with $H_{spin}$.


\begin{figure*}
\includegraphics[width=5.5in]{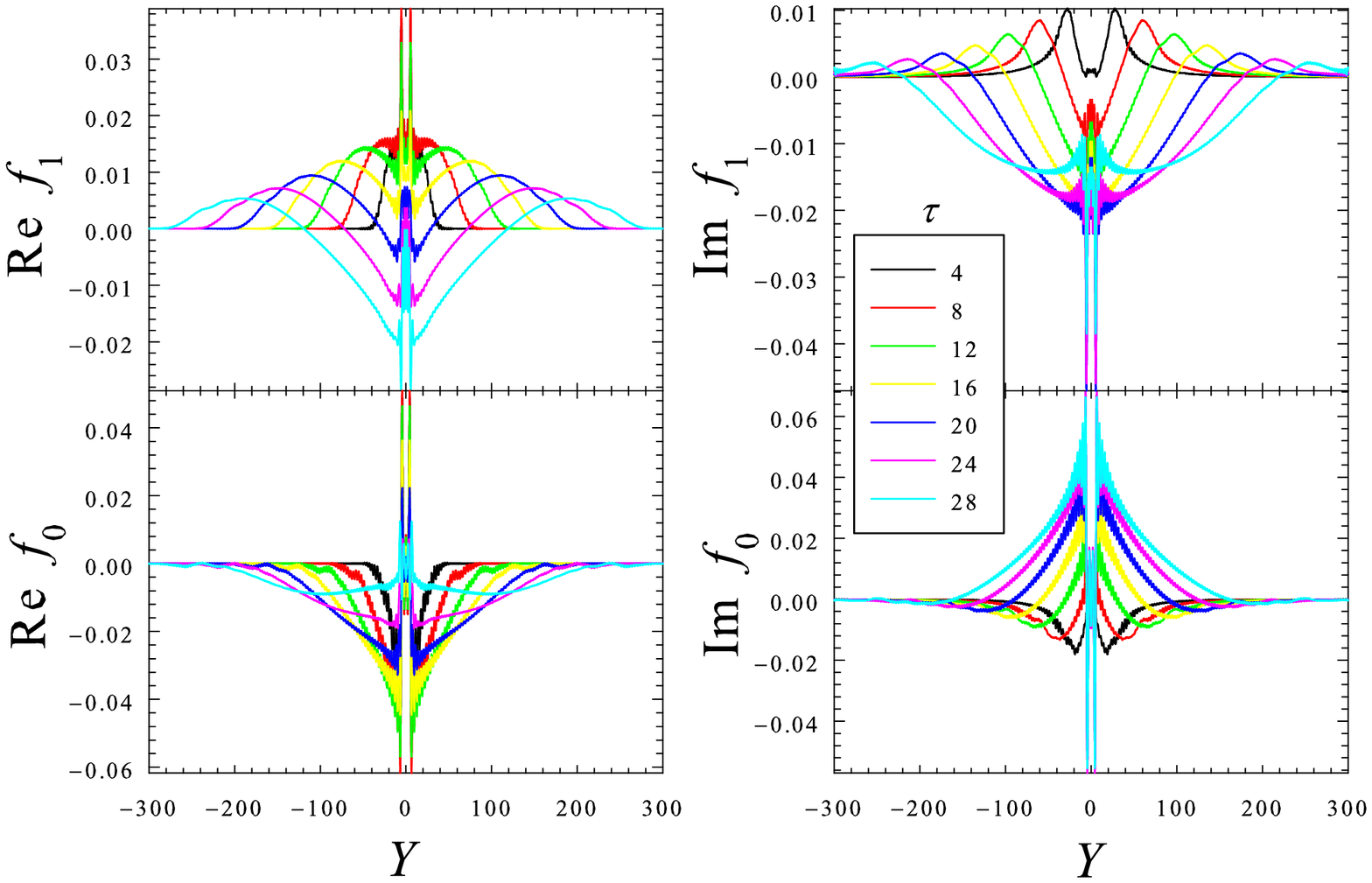}
\includegraphics[width=5.5in]{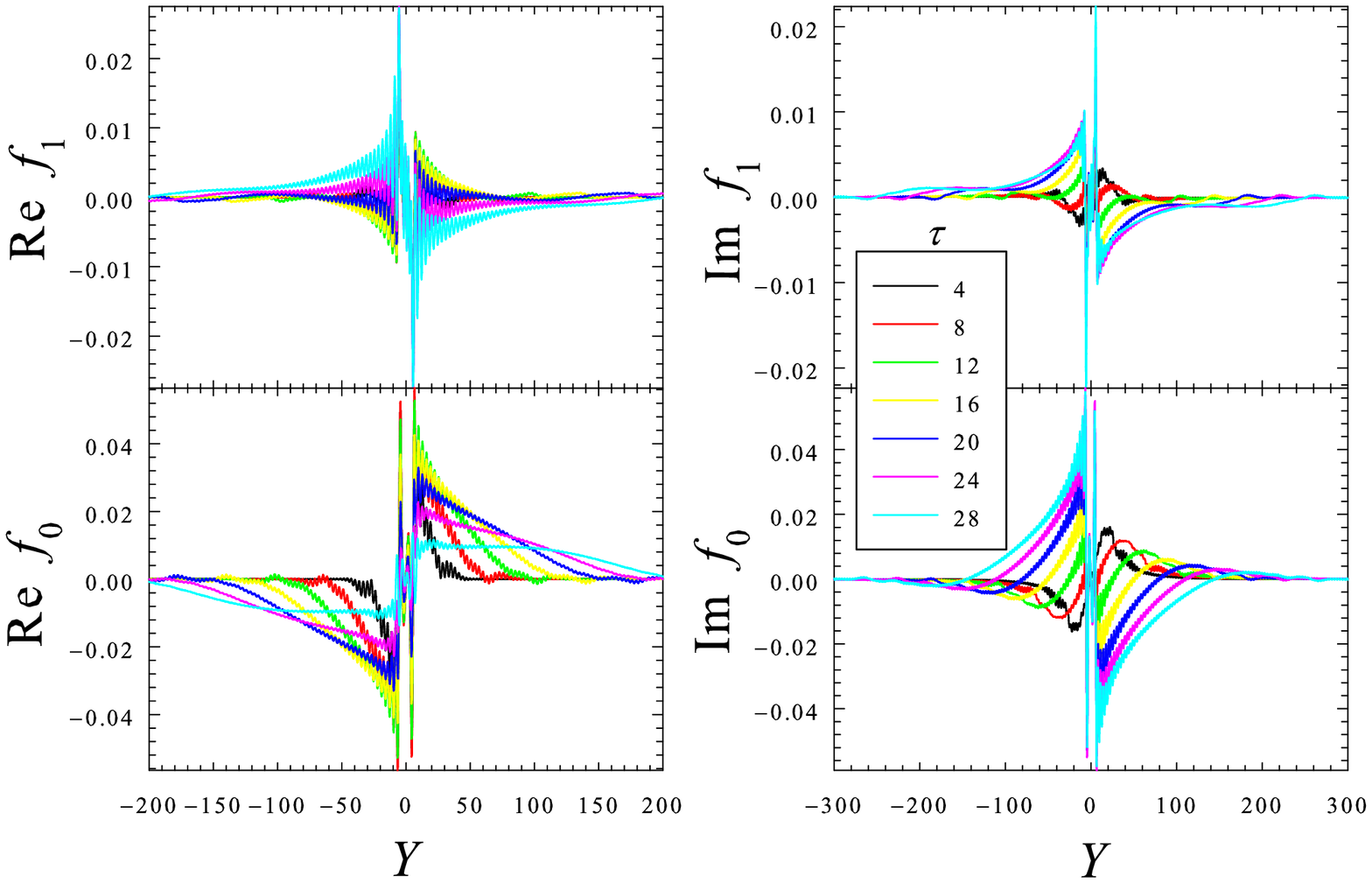}
\caption{ (Color online)
The temporal dependence of the 
$0$ and $\pi$ state junction triplet amplitudes. 
They are arranged as in Fig.~\ref{fig2} and 
plotted as a function
of  $Y$
for several equally spaced values of $\tau$. These results are at a fixed
value of the spin flip parameter, $H_{spin}=0.2$,
where both the 
$\pi$ state  and  the $0$ state are stable (see Fig.~\ref{fig1}). } 
\label{fig3}
\end{figure*}
In Fig.~\ref{fig3} we show results for the triplet amplitudes in the same format
as in Fig.~\ref{fig2} but at fixed $H_{spin}$ 
and several values of $\tau$,
so that the explicit time evolution of the different-time triplet states 
can be visualized.
As before, 
the real or imaginary part of $f_0$ or $f_1$ are
appropriately labeled whether discussing the
$0$ or $\pi$ state configurations (top four or bottom four panels respectively).
All results are for an intermediate spin flip transparency  $H_{spin}=0.2$, 
where both junction states have
very approximately the same condensation energy 
(see Fig.~\ref{fig1}). The region of the sample shown here is wider than
that in Fig.~\ref{fig2} because the spatial range over which the correlations
extend is now wider,  at 
larger times. We see than the triplet correlations, which of course
vanish identically at $\tau=0$ already pervade the $S$ layer at the earliest
times shown. 

The real part of the  amplitude $f_1$ has, in the $0$-state, a maximum 
in the S region 
that keeps propagating outwards as
$\tau$ increases, reflecting the longer penetration of the correlations into S.
This maximum becomes shallower with increasing $\tau$ however.
This increased penetration occurs also for both the real and imaginary components of $f_0$, 
although in the latter case the 
maximum value of the amplitude in S  occurs near the interface 
except in some instances
at the
longest times studied. The largest $\tau$ 
studied was determined by the 
need to avoid finite size effects: after the different-time correlations 
have pervaded the
entire S portion of the sample the results would be contaminated 
by outer boundary effects. 
Referring still to the $0$ state, we again have the situation where 
for a given time $\tau$ and fixed value of $H_{spin}$, the 
${\rm Im} f_1$ 
tends to have pervaded more of the superconductor region
than the real part.   
The $\pi$ amplitudes have some similarities with those in Fig.~\ref{fig2}, in that 
${\rm Re }f_1$ 
is smaller and has a weak dependence on the scattering strength. 
Although ${\rm Im} f_1$ demonstrates a stronger  dependence on $H_{spin}$,
and besides prominently peaking at the interfaces,
the overall magnitudes for both components are effectively reduced 
for all values of $\tau$ shown.

\begin{figure}
\includegraphics[width=3.5in]{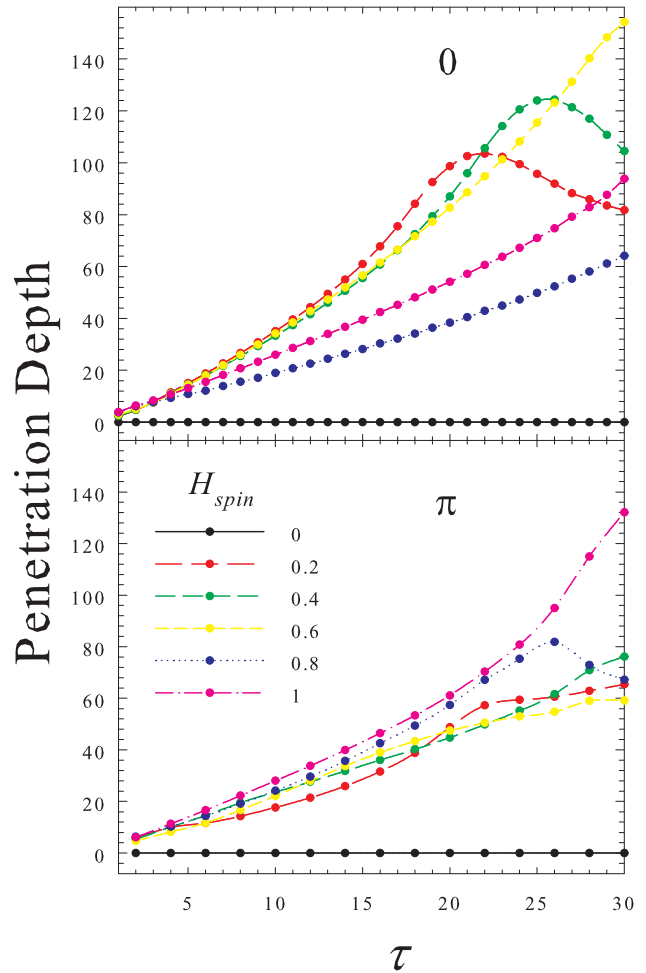}
\caption{(Color online)  
The penetration depth, as defined in dimensionless units 
in Eq.~(\ref{pl}),
of the equal spin $f_1$ triplet component into the S material, plotted as a
function of $\tau$ for both the ``0" and ``$\pi$" configurations (as labeled).}
\label{fig4}
\end{figure}
This penetration of the triplet correlations into the S material can be 
conveniently described in terms of 
time-dependent penetration depths. These can be calculated
for either $f_0$ or $f_1$. We will focus here on 
the real parts of $f_1$, with
the understanding that an analogous approach could be followed
for either $f_0$ or the imaginary components.
However, it is also evident from Fig.~\ref{fig2} that for $f_0$,
the penetration of triplet correlations into S is only weakly dependent 
on $H_{spin}$, which is consistent with the fact that $f_0$ does not emerge
from $H_{spin}$ only. 
The method to extract any sort of characteristic length depends of 
course on the
problem at hand, and for the $f_1$ amplitudes we 
find the following definition yields sensible results,
\begin{equation}
\ell(\tau)=
\frac{\int_{S} dY |Y-Y_0| |{\rm Re} \lbrace f_1(Y,\tau) \rbrace| }{\int_{S} dY |{\rm Re} \lbrace f_1(Y,\tau) \rbrace|},
\label{pl}
\end{equation}  
which is slightly modified from that used 
previously.\cite{hvb} The definition 
Eq.~(\ref{pl})  accounts better for cases where the overall shape of the
amplitudes (see Figs.~\ref{fig2} and \ref{fig3}) 
varies depending on the parameter values. 
The coordinate shift, $Y-Y_0$, accounts for measuring the  distance from 
the interface: for our coordinates, $Y_0 = D_F/2$  
and the integration extends over the
S region. 
This definition gives the expected result
if the function $f_1$ were a pure decaying exponential.
The values of the dimensionless $\ell$ are in units of $k_F^{-1}$.
The results are plotted in Fig.~\ref{fig4} as a function of $\tau$.
These results are for 
the same range of $H_{spin}$ as in the previous figures,
as shown in the legend.
Both
the $0$ (top) and $\pi$ (bottom) state penetration depths are shown.
There is an approximately linear behavior, in both $0$ and $\pi$ cases, 
at earlier times 
and a deviation from linearity at later times.
For the $0$ state, the values of $\ell$ reach discernible 
maxima that get shifted to larger times
with increased spin-dependent scattering rates. For 
larger values of 
$H_{spin}$,
the times necessary to reach the peak would presumably correspond 
to times where these triplet correlations would
have reached the boundary and
finite size effects would be a concern. 
This increase for  larger $H_{spin}$, however,
should not continue much beyond the range shown, since we know\cite{hvb} that
the triplet correlations do eventually begin to decay after a given 
characteristic time $\tau$ of a few times $2\pi$. 
It is clear, however, that the penetration extends over a wide range of 
times over regions much larger than the superconducting coherence
length.
The results here are  consistent with what we saw in Fig.~\ref{fig2},
where at $\tau=20$, we see for $H_{spin} = 0.2,0.4$, and $0.6$,
the manifestation of secondary broad maxima at large Y, and little variation among the $f_1$ amplitudes, 
while the  other values (0.8 and 1) have generally a monotonic decline, and thus for these higher values they have 
smaller characteristic penetration depths.
The penetration of equal spin triplet correlations into a $\pi$ junction 
is much less dependent on the spin scattering strength 
than for $0$ junctions except of course at very
small $H_{spin}$. 
This is again consistent with what was shown
in Fig.~\ref{fig2}, where the real parts of the $f_1$ correlations
demonstrated  very little dependence on $H_{spin}$, and were weaker overall,
away from the interface, than their  $0$ junction counterparts.
This is reflected in the penetration depth behavior, where the 
depth is reduced compared to the $\pi$ case and the penetration is
similar for the broad range of spin dependent scattering strength. 
The only exception
being the extreme case of $H_{spin}=1$, which always has a greater penetration,
but this too agrees well with what is observed in Fig.~\ref{fig2}.

\begin{figure*}
\includegraphics[width=6in]{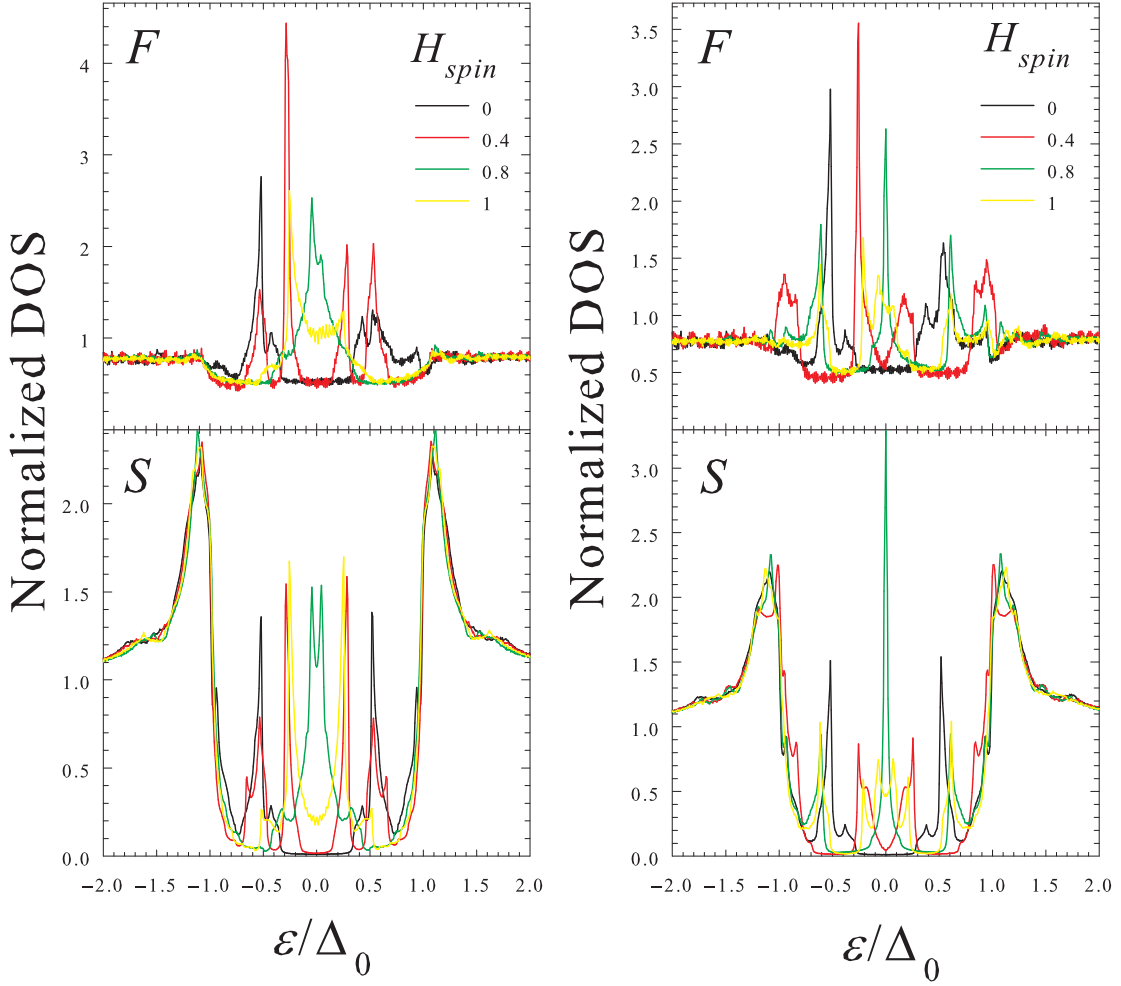}
\caption{ (Color online)
The local density of states, normalized to unity in the normal state of the S
material. The top panels show results within
the thin $F$ layer, while
the bottom one  is for the superconductor. The left column is for the 0
state and the right one for the $\pi$ state.}
\label{fig5}
\end{figure*}  
The density of states, measured in principle by in STM experiments, 
is one way of probing indirect evidence of the triplet
superconducting states, and carries valuable energy resolved spectroscopic 
information.
We therefore show next, in Fig.~\ref{fig5}, the DOS, computed from Eq.~(\ref{dos}). 
For computational purposes, we represent the delta functions in
Eq.~(\ref{dos}) as 
the low $T$ limit of the derivative of the corresponding Fermi functions.
We find that a fine mesh of $\varepsilon_\perp$ is necessary to properly 
calculate the energy resolved DOS, and 
that there are significant contributions to the DOS from both
longitudinal propagation (small $\varepsilon_\perp$) and 
from large off-normal incidence at the interface (large $\varepsilon_\perp$).
The junction here is therefore  not appropriately described within 
the tunneling limit, \cite{linder} and would yield differing results
if only a narrow tunneling cone was used in the calculation.
By considering spin active interfaces, it was  shown\cite{cottet} that various 
signatures arise in the DOS, 
including a unique double gap structure. 
In Fig.~\ref{fig5} the DOS is shown,
normalized to its bulk value in S, 
as a function of energy $\epsilon$ (relative
to $E_F$), in units
of the bulk $\Delta_0$. 
We consider
four  values of 
$H_{spin}$ for both the $0$ state (left panels) and the $\pi$ state 
(right panels). 
The top panels shows results summed over both spins 
and averaged over the entire thin $F$ layer 
(label ``F") 
while the bottom 
panels
shows results (also summed over spins) 
averaged over  the whole length of one of the superconductors (label ``S").
For the S regions, in both the $0$ and $\pi$ junctions
there are BCS-type peaks at
$\epsilon/\Delta_0 \approx \pm 1$, reflecting the bulk-like behavior. Inside the
region of the bulk gap there is a secondary structure reflecting Andreev states.
These secondary peaks were also found in Ref.~\onlinecite{us69}. 
We also see that the subgap  peaks arise even in the absence
of interface spin activity (at $|\epsilon/\Delta_0| \approx 0.5$),
and originate mainly from individual spin channels, depending on the sign of 
the energy:
the prominent subgap peak at negative energy is due to the occupation of spin-up
quasiparticles, $N_\uparrow$, while its positive energy counterpart arises 
from $N_\downarrow$.
The position of these subgap peaks varies with $H_{spin}$ in a way
that seems to reflect the condensation energy in Fig.~\ref{fig1}, particularly
in the 0 state. 
At higher energies, $|\epsilon/\Delta_0|>1$, both spin bands contribute equally 
to the DOS.
In general within S and for the range of energies shown, the
approximate  relation, 
$N_{\uparrow}(\epsilon) \approx N_{\downarrow}(-\epsilon)$, holds,
giving the observed symmetry in energy for the total DOS in the S region 
(Fig.~\ref{fig5} bottom panels).
Increasing $H_{spin}$, tends to flip the spins at the interface, and thus bound states
in S predominantly occupied by a given spin species, become replaced by the opposite spin quasiparticles.
This is confirmed by examination of the individual 
spin density of states (Eq.~\ref{dos}).
The half metallic ferromagnet 
modifies this bound state picture in that region
due to the existence of only one spin band at the Fermi level (top panels).
Here the majority of the  energy-resolved
states must come from spin-up quasiparticles. The spin-flip processes inherent to the scattering events at
the interface (the parameters $H_{spin}$ or $I$) in conjunction with 
proximity effects can however 
cause an enhancement of subgap bound states at small energies
attributed to a small number of minority spin states in the magnet.
Also due to the strong
spin splitting in that region, there is significant particle-hole asymmetry. 
At higher values of 
$|\epsilon/\Delta_0|$  the correct limit is approached 
($\approx (1/2)(1+I)^{1/2}$). The structure in the region 
$|\epsilon/\Delta_0| < 1$ 
is now considerably more complicated but consistent with
what is seen in the S region: the number of states at zero energy is greater when the 
condensation  energy is very small (see Fig.~\ref{fig1}).  

We turn now to the trends the peaks in S follow as a function of $H_{spin}$,  
for both junction configurations.
As $H_{spin}$ is increased from $0$ to about $0.8$, the value at which the condensation free energies 
are near their minimum
values (see Fig.~\ref{fig1}), we see that these peaks tend to merge and there is a
zero-energy single or double peak signature, depending on whether it is a $\pi$ or $0$ junction 
respectively.   
For  $H_{spin}=1$, the peaks widen, to nearly the same energies
as for $H_{spin}=0.4$. It does appear that for both junction states, depending on the 
spin strengths, there exists a small subgap region which resembles an energy gap in the DOS.
Strictly speaking, however, the energy spectrum is gapless 
for the whole range from no spin flipping to strong activity at the interface;
there is always a finite, albeit small in some cases, number of states within this
Andreev bound state region. Adding a sufficiently strong spin-independent scattering or 
further increasing $d_S$ would eventually create a region with no states.
The existence of the half metal and the interface scattering
thus still influences the superconductor when considering spatially averaged behavior. 

\begin{figure}
\includegraphics[width=3.4in]{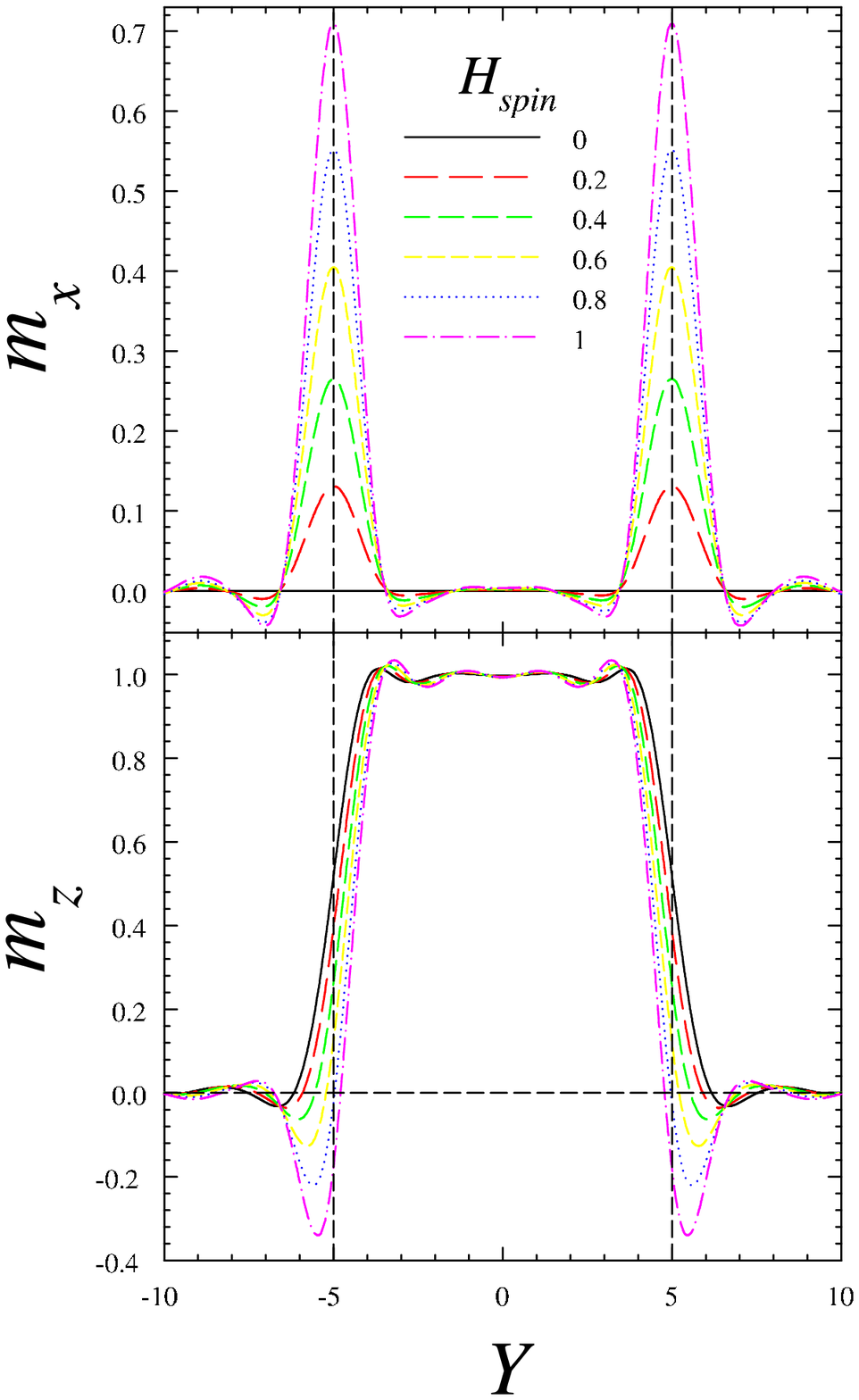}
\caption{ (Color online)
The $x$  (top panel) and $z$ components (bottom panel) of the local magnetic
moment normalized by $\mu_B$ (see Eq.~(\ref{mm})) plotted as a 
function of the dimensionless coordinate $Y$ for 
$H_{spin}=$ from 0 to 1 at 0.2 intervals. Only  a limited range of $Y$ is
included. The $0$ and $\pi$ state results are identical for the entire range of spin
flipping considered here.}
\label{fig6}
\end{figure}  
To gain further insight into the relative 
proximity effects inherent to these junctions, we present in Fig.~\ref{fig6}  
results for the influence of $H_{spin}$ on the
inverse proximity effect, that is, on the local magnetic moment 
as defined in Eq.~(\ref{mm}).
We find that  
if the large $\varepsilon_\perp$  off-normal trajectories are not fully included, 
the magnetic moment does not reach its proper normalized values  and spatial properties.
The calculation of $m_z$ and $m_x$ thus serves as another check to ensure that all the 
requisite states are included for other calculations. 
The results for the local magnetization
(as defined above in Eq.~(\ref{mm}) and normalized by $\mu_B$) 
are shown in Fig.~\ref{fig6}. They display the 
penetration of the magnetization into the superconductor region, as well as the
weakening spin polarization in F. This is currently a
topic of extreme interest, experimentally and theoretically, especially when trying to clarify the 
complicated spin structure in these systems.
One can also view the introduction of
spin scattering as an ultra narrow domain wall at the interfaces.
In the figure both the $x$ and the $z$
components of the local magnetic moment are plotted as a function of $Y$ for
several values of $H_{spin}$ in the range $0 \le H_{spin} \le 1$. 
The results are the same for 0 and $\pi$ states. Results for the 0 state are shown. 
The $x$ component vanishes by symmetry 
since the exchange field lies in the $z$ direction 
when
$H_{spin}=0$. At nonzero values of this parameter, $m_x$ grows very quickly in the
interface region, while remaining zero in the center of the $F$ layer and
also, of course, deep in the superconductor. For this reason only the central
part of the system is included in the plot. The $z$ component (bottom panel) has
a very weak variation with $H_{spin}$ near the center, $Y=0$, of the half metal, but it does 
show a dependence on the scattering strength
in the superconductor near
the interface. The inverse proximity effect is clearly evident where
the induced magnetization component $m_z$ in the  S region near the interface is oppositely directed 
to that in F, 
effectively screening the magnetization in F, by an amount that increases monotonically with $H_{spin}$.
On the ferromagnet side very near the interface, this component of the magnetic moment 
correspondingly weakens with increased $H_{spin}$, before rising up to
near the half metallic bulk value of unity. 
The screening effect in the superconductor is apparently stronger for $m_z$ than $m_x$. 
For the latter 
the induced magnetic moment is very similar in adjacent regions near the boundaries,
and  there is  near symmetry about the interfaces. 
The  component $m_x$ reverses signs in both F and S for all
$H_{spin}$, while $m_z$ is briefly negative for only the larger $H_{spin}$,
demonstrating the competing effects from the exchange field and spin scattering strength.
The observed spatial characteristic of each component in Fig.~\ref{fig6}
reveal that ${\bf m}$ in the vicinity of
the interfaces 
tends to not only change magnitude as a function of position,
but it also {\it rotates}.
The magnetization also changes direction
as a function of $H_{spin}$
for fixed $Y$, illustrating again the important
role the proximity effects play on the relevant self-consistent 
quasiparticle wavefunctions and energies.

\section{conclusions} 
\label{conclusions} 
We have investigated the effect of interfacial spin-flip scattering on the
triplet correlations that emerge in an SFS trilayer. We have studied this system
by solving in a fully self-consistent way the BdG equations in the clean 
limit. We have considered both 0 and $\pi$ junctions and found that
the results depend strongly on the junction state. 
Triplet amplitudes, odd in time as  required by the  
Pauli principle, have been found to exist 
and we have studied them in detail for the case where F is half-metallic. 
We have found that the $m=\pm 1$ triplet amplitudes emerge and then subsequently increase (at finite times) very rapidly
with the dimensionless spin flip parameter $H_{spin}$.
The degree at which the equal-spin triplet correlations pervade the S layers has been discussed 
in connection with the respective penetration depths. 
We also have presented results for the local energy resolved DOS averaged over both the S and F regions 
as a function of the spin-flip rate. The $0$ or $\pi$ state signatures may provide clues as to  
how different-time triplet states indirectly 
influence the subgap energy spectrum.
We have also considered
the inverse proximity effect (the penetration of the magnetization into S and its weakening
in F) and found that near the interfaces the magnetization rotates as a function of 
position or of $H_{spin}$. Ultimately, the induced spin imbalance in the superconductor   effectively 
screens the polarizing effects of the half metal.

\begin{acknowledgments}
This project was supported in part by 
a grant of supercomputer resources provided by the DoD 
High Performance Computing Modernization Program (HPCMP) and 
NAVAIR's ILIR program
sponsored by ONR.
We  thank  I. Krivorotov for useful discussions.
\end{acknowledgments}


\end{document}